\documentclass{article}
\usepackage{spconf,amsmath,epsfig}

\usepackage{booktabs}
\usepackage{url}
\usepackage{xcolor}
\usepackage{graphicx}

\title{Prediction of Sentinel-2 multi-band imagery with attention BiLSTM for continuous earth surface monitoring}
\threeauthors
  {Weiying Zhao}
  {Deep Planet \\London, UK }
  {Natalia Efremova}
  {Queen Mary University and \\the Alan Turing Institute,	London, UK}
    {}
  {\\	}

\begin{document}
%
\maketitle
%

\begin{abstract}

Continuous monitoring of crops and forecasting crop conditions through time series analysis is crucial for effective agricultural management.  This study proposes a framework based on an attention Bidirectional Long Short-Term Memory (BiLSTM) network for predicting multiband images. Our model can forecast target images on user-defined dates, including future dates and periods characterized by persistent cloud cover. By focusing on short sequences within a sequence-to-one forecasting framework, the model leverages advanced attention mechanisms to enhance prediction accuracy.  Our experimental results demonstrate the model's superior performance in predicting NDVI, multiple vegetation indices, and all Sentinel-2 bands, highlighting its potential for improving remote sensing data continuity and reliability.

\end{abstract}


\section{Introduction}

Accurately predicting missing bands in remote sensing data is critical for continuous monitoring and analysis of Earth's surface processes. Traditional time series interpolation methods are commonly used for reconstructing historical missing images. However, these methods often struggle with data quality issues, such as cloud cover, that can obscure critical data, especially in optical satellite sensors \cite{gerber2018predicting}. Recent studies also utilize Synthetic Aperture Radar (SAR) signals, which are less affected by cloud cover, to impute missing images due to their robustness in adverse weather conditions \cite{mandal2019investigation, zhao2023combining, rossberg2024dense}. The Gaussian process has also been recognized as a potent method for imputing bands in time series data by capturing the spatial coherence and temporal regularity inherent in remote sensing data \cite{pipia2019fusing}.

The integration of attention mechanisms in LSTM networks has shown promising results in various studies \cite{chen2022joint, zhang2022watershed, miller2024deep}. These mechanisms focus the model on relevant parts of the input data, improving the accuracy of predictions in complex datasets characterized by missing or incomplete information \cite{wang2023comprehensive, wang2023dafa}. For instance, Gerber et al. (2018) \cite{gerber2018predicting} highlighted the potential of spatio-temporal prediction methods to fill gaps in datasets with extensive missing values due to cloud cover and other artefacts.

Our proposed framework utilizes a BiLSTM network with an attention mechanism, enhancing the model's ability to handle short sequences in a sequence-to-one forecasting framework. This approach builds on the demonstrated effectiveness of LSTM networks in handling temporal data gaps by learning from both past and future contexts, which is particularly advantageous for predicting cloud-free images acquired in user-defined time.

\section{Method}

\begin{figure*}[h]
    \centering
    \includegraphics[width=1.6\columnwidth]{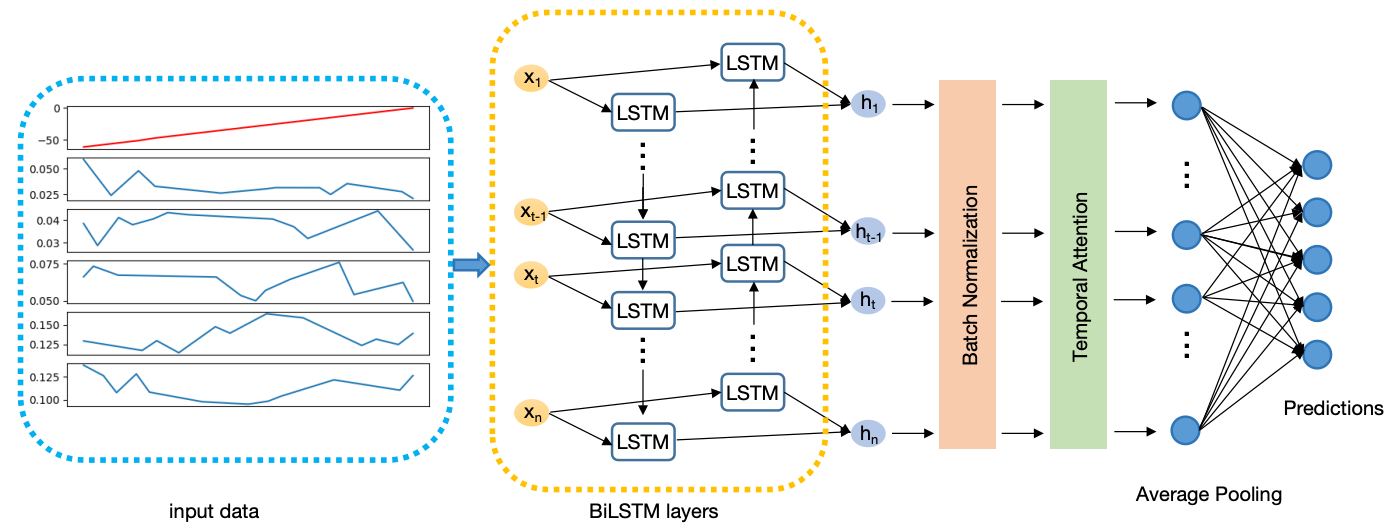}
	\caption{Flowchart of the vanilla attention BiLSTM. The processing flow in this figure takes 5 time series bands as an example. The additional times series, which is highlighted in red, is the time difference time series.}
	\label{fig:attention_BiLSTM}
\end{figure*}

In designing and implementing our neural network architecture (Fig.\ref{fig:attention_BiLSTM}), we integrated advanced deep learning techniques to address complex sequence learning tasks. The model commences with an input layer configured to accommodate the dimensions of the input sequence, specifically tuned to the time steps and feature counts of the dataset in use. This is followed by a series of LSTM layers, where the initial layer is bidirectional, enabling the network to learn from the sequence data in both forward and reverse directions. Each LSTM layer is accompanied by batch normalization and ReLU activation functions, which are critical for stabilizing the neural network by normalizing the activations and introducing non-linearity, respectively.

Furthermore, we incorporate an attention mechanism, which selectively weighs the importance of different time steps across the input sequence. This is achieved through a dense layer with softmax activation that outputs attention probabilities. These probabilities are then element-wise multiplied with the LSTM outputs to emphasize features of higher relevance to the task at hand.
Post-attention application, the model leverages a Global Average Pooling 1D layer to condense the feature map across time steps into a single vector, reducing the model's complexity and computational demands while retaining essential temporal features. The output layer, constructed with a dense layer, is tailored with an adjustable number of neurons corresponding to the desired outputs and utilizes a linear activation function to produce the final prediction.


The methodological choices in constructing this neural network, from the layered architecture to the incorporation of attention and advanced regularization techniques, are designed to optimize performance for predicting multi-dimensional outputs from sequential data. This comprehensive approach not only ensures robust learning capabilities but also adapts effectively to various types of sequence-based prediction tasks.

\begin{figure}[h]
    \centering
    \includegraphics[width=.9\columnwidth]{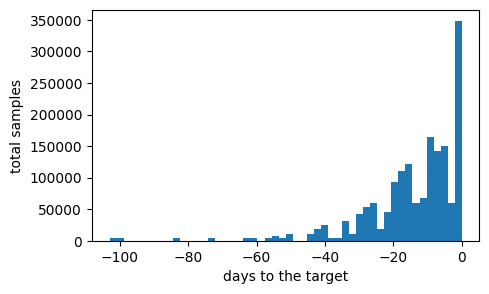}
	\caption{Time difference distribution. The time difference is between the target image acquisition time and the previous 5 image acquisition time.}
	\label{fig:time_differences}
\end{figure}

\section{Experimental results and discussion}
\subsection{Dataset}
\begin{figure*}[t]
    \centering
    {
        \includegraphics[width=.9\textwidth]{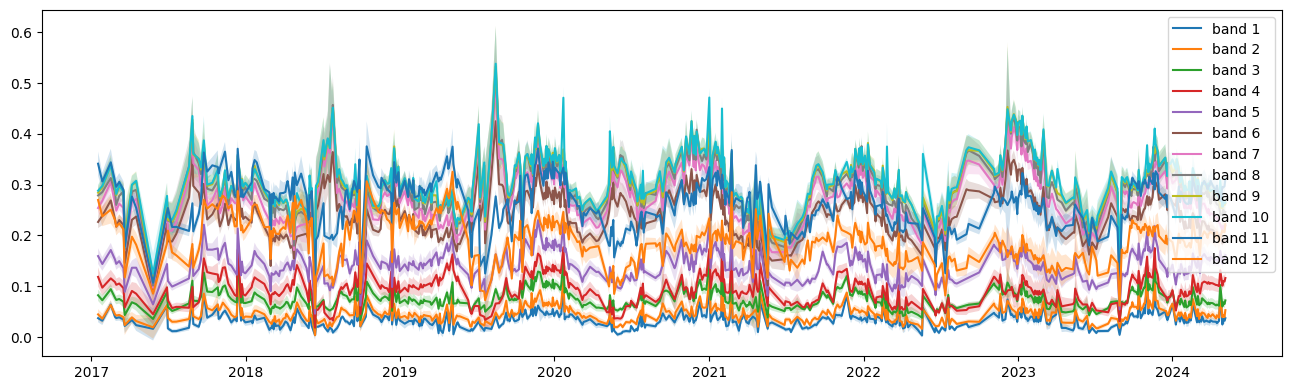}
        \label{fig:first_row}
    }
    {
        \includegraphics[width=.9\textwidth]{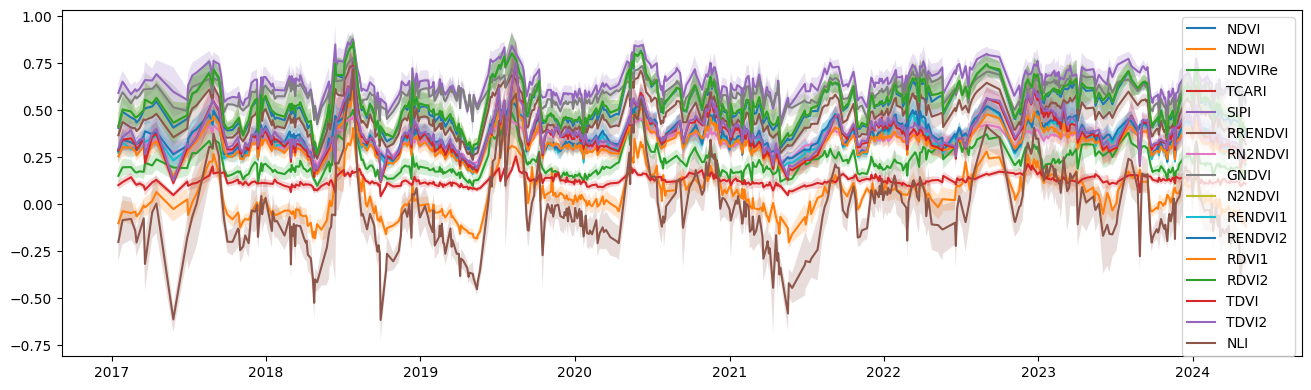}
        \label{fig:second_row}
    }
    \caption{Multibands time series mean and standard deviation across different bands. (up) Original time series. (bottom) Time series vegetation indices. Most of the cloud-affecting Sentinel-2 images have been removed based on the NDSI band.}
    \label{fig:multibands_time_series}
\end{figure*}

To compare the proposed method with other LSTM-based methods, we tested them with 3 databases prepared using selected multitemporal cloud-free Sentinel-2 images. Only the areas acquired over vineyards are used to prepare the database. 
The time steps are 5, and the channels all contain the time difference channel. The time difference (Fig.\ref{fig:time_differences}) is computed between the target acquisition time and the previous 5 band acquisition times. It can be defined by the user accounting for the specific requirement. The mean and standard deviation of different Sentinel-2 bands and vegetation indices computed with the images acquired in a vineyard are introduced in Fig.\ref{fig:multibands_time_series}.

\begin{figure}[h]
    \centering
    \includegraphics[width=.85\columnwidth]{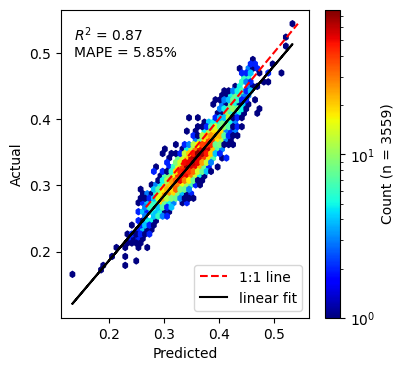}
	\caption{Scatter plot of predicted and real NDVI values. The predicted image is unseen during the model preparation.}
	\label{fig:ndvi_forecasting}
\end{figure}

\begin{table*}
\renewcommand{\arraystretch}{1.2}
\begin{center}
\begin{tabular}{c || c | c | c | c | c | c} 
\hline\hline
Methods & RMSE-NDVI & MAPE-NDVI & RMSE-VIs & MAPE-VIs & RMSE-s2 & MAPE-S2 \\
\hline\hline
attention BiLSTM & 0.0323 & \underline{4.796} & \textbf{0.0281} & \textbf{5.348} & \underline{0.0174} & \textbf{6.690} \\
attention LSTM & \textbf{0.0315} & \textbf{4.469} & 0.0313 & 5.545 & \textbf{0.0169} & 7.403 \\
BiLSTM & 0.0329 & 5.174 & {0.0318} & {5.944} & 0.0183 & 7.3700 \\
attention ConvLSTM2D & 0.0333 & 5.2096 & \underline{0.0302} & 5.576 & 0.0176 & 7.991 \\
ConvLSTM2D & \underline{0.0320} & 4.854 & 0.0313 & \underline{5.474} & 0.0179 & \underline{7.235} \\
\hline\hline
\end{tabular}
\caption{Performance comparison of different methods. The best
results are in bold and the second best are underlined}
\label{tab:forecasting_results}
\end{center}
\end{table*}

\subsection{NDVI forecasting}

The performance metrics in NDVI forecasting, as shown in Tab.\ref{tab:forecasting_results}, reveal that the attention LSTM model achieved the best performance with the lowest RMSE (0.0315) and MAPE (4.469), indicating a robust capability to model and predict NDVI accurately. This outcome suggests that LSTM with attention mechanisms effectively captures temporal dynamics crucial for NDVI estimation. The attention BiLSTM provides the second best MAPE result, and one test result over unseen data can be seen in Fig.\ref{fig:ndvi_forecasting}. The ConvLSTM2D model, despite not using an attention mechanism, also performed commendably, achieving the second-best RMSE (0.0320) and a competitive MAPE (4.854). This section underscores the importance of both LSTM-based models and attention approaches in handling the temporal-spatial complexity of NDVI data from satellite imagery.

\begin{figure*}
    \centering
    \includegraphics[width=.8\textwidth]{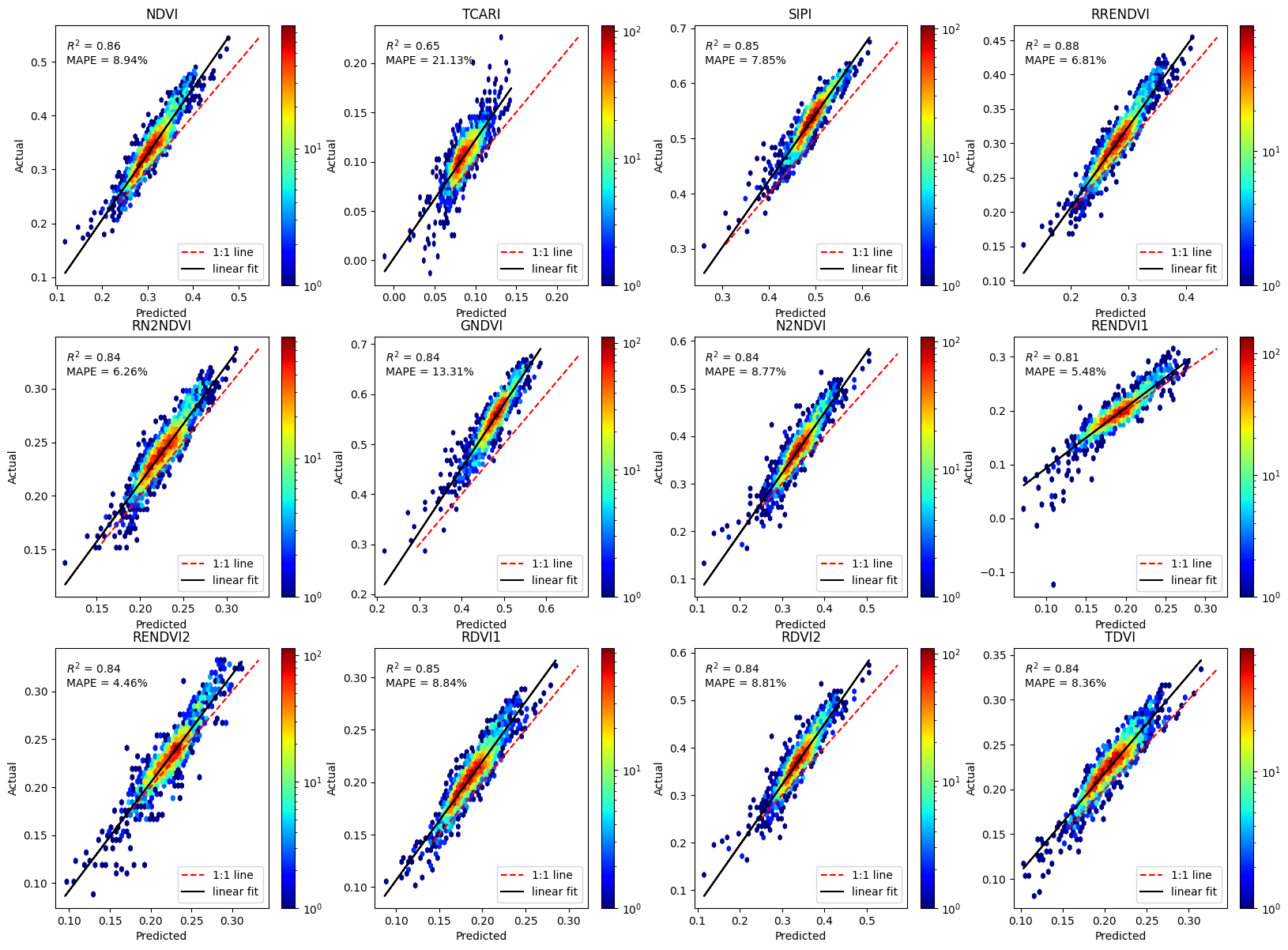}
	\caption{Scatter plot of predicted and real vegetation indices.}
	\label{fig:multi_VIs_forecasting}
\end{figure*}

\subsection{Multi vegetation indices forecasting}
In forecasting multiple vegetation indices, the attention BiLSTM model outperformed other architectures, achieving the best RMSE (0.0281) and MAPE (5.348). One forecasting example can be found in Fig.\ref{fig:multi_VIs_forecasting}. These results suggest that the bidirectional processing of LSTM layers, combined with an attention mechanism, provides a significant advantage in handling the complexities of multiple vegetation indices.
Notably, the attention ConvLSTM2D model also showed strong performance with the second-best RMSE (0.0302), indicating the effectiveness of integrating convolutional layers for spatial feature extraction alongside LSTM's temporal analysis.

\subsection{Sentinel-2 bands forecasting}

For the Sentinel-2 bands forecasting, the attention BiLSTM model again stands out, recording the best MAPE (6.690) and a highly competitive RMSE (0.0174), underscoring its effectiveness in processing multi-spectral satellite data. The attention LSTM model provided the lowest RMSE (0.0169), although with a slightly higher MAPE, highlighting the trade-offs between different error metrics and the model's sensitivity to outlier predictions. From Fig.\ref{fig:multibands_time_series}, we can see that Sentinel-2 time series bands have larger standard deviations than Sentinel-2 vegetation indices. It highlighted the robustness of the model performances.
These comparisons show how different architectures manage the high-dimensional nature of Sentinel-2 data, with attention mechanisms improving the model's focus on relevant spectral bands.

\section{Conclusion}


This study successfully demonstrated the application of an attention BiLSTM network for the prediction of multiband remote sensing images, aimed at enhancing crop monitoring and forecasting under varying cloud cover conditions.
The experimental results underscore the model's potential to provide reliable data continuity in remote sensing, particularly in regions and seasons where cloud cover frequently obscures the earth's surface. The ability of our model to forecast images on user-defined dates, including during adverse weather conditions, represents a substantial advancement in the field of agricultural management and environmental monitoring.

Future research will aim to extend this work by focusing on the prediction of missing multiband images over entire seasons, further enhancing the model's ability to handle diverse and complex environmental conditions.

\small
\bibliographystyle{IEEEbib}
\bibliography{refs}

\begin{thebibliography}{10}

\bibitem{gerber2018predicting}
Florian Gerber, Rogier de~Jong, Michael~E Schaepman, Gabriela Schaepman-Strub, and Reinhard Furrer,
\newblock ``Predicting missing values in spatio-temporal remote sensing data,''
\newblock {\em IEEE Transactions on Geoscience and Remote Sensing}, vol. 56, no. 5, pp. 2841--2853, 2018.

\bibitem{mandal2019investigation}
Dipankar Mandal, Mehdi Hosseini, Heather McNairn, Vineet Kumar, Avik Bhattacharya, YS~Rao, Scott Mitchell, Laura~Dingle Robertson, Andrew Davidson, and Katarzyna Dabrowska-Zielinska,
\newblock ``An investigation of inversion methodologies to retrieve the leaf area index of corn from c-band sar data,''
\newblock {\em International Journal of Applied Earth Observation and Geoinformation}, vol. 82, pp. 101893, 2019.

\bibitem{zhao2023combining}
W~Zhao, F~Yin, H~Ma, Q~Wu, J~Gomez-Dans, and P~Lewis,
\newblock ``Combining multitemporal optical and sar data for lai imputation with bilstm network,''
\newblock {\em arXiv preprint arXiv:2307.07434}, 2023.

\bibitem{rossberg2024dense}
Thomas Ro{\ss}berg and Michael Schmitt,
\newblock ``Dense ndvi time series by fusion of optical and sar-derived data,''
\newblock {\em IEEE Journal of Selected Topics in Applied Earth Observations and Remote Sensing}, 2024.

\bibitem{pipia2019fusing}
Luca Pipia, Jordi Mu{\~n}oz-Mar{\'\i}, Eatidal Amin, Santiago Belda, Gustau Camps-Valls, and Jochem Verrelst,
\newblock ``Fusing optical and sar time series for lai gap filling with multioutput gaussian processes,''
\newblock {\em Remote Sensing of Environment}, vol. 235, pp. 111452, 2019.

\bibitem{chen2022joint}
Baili Chen, Hongwei Zheng, Lili Wang, Olaf Hellwich, Chunbo Chen, Liao Yang, Tie Liu, Geping Luo, Anming Bao, and Xi~Chen,
\newblock ``A joint learning im-bilstm model for incomplete time-series sentinel-2a data imputation and crop classification,''
\newblock {\em International Journal of Applied Earth Observation and Geoinformation}, vol. 108, pp. 102762, 2022.

\bibitem{zhang2022watershed}
Qiang Zhang, Ruiqi Wang, Ying Qi, and Fei Wen,
\newblock ``A watershed water quality prediction model based on attention mechanism and bi-lstm,''
\newblock {\em Environmental Science and Pollution Research}, vol. 29, no. 50, pp. 75664--75680, 2022.

\bibitem{miller2024deep}
Lynn Miller, Charlotte Pelletier, and Geoffrey~I Webb,
\newblock ``Deep learning for satellite image time-series analysis: A review,''
\newblock {\em IEEE Geoscience and Remote Sensing Magazine}, 2024.

\bibitem{wang2023comprehensive}
Qunming Wang, Yijie Tang, Yong Ge, Huan Xie, Xiaohua Tong, and Peter~M Atkinson,
\newblock ``A comprehensive review of spatial-temporal-spectral information reconstruction techniques,''
\newblock {\em Science of Remote Sensing}, p. 100102, 2023.

\bibitem{wang2023dafa}
Heshan Wang, Yiping Zhang, Jing Liang, and Lili Liu,
\newblock ``Dafa-bilstm: Deep autoregression feature augmented bidirectional lstm network for time series prediction,''
\newblock {\em Neural Networks}, vol. 157, pp. 240--256, 2023.

\end{thebibliography}

\end{document}